\documentstyle[11pt,newpasp,twoside,epsf]{article}
\def\etal{{\it et al.~}}

\begin{document}

\title{1997 October Event(s) in the Crab Pulsar}
\author{D. C. Backer}
\affil{Astronomy Department, University of California, Berkeley, CA, USA}

\begin{abstract}
In October 1997 daily monitoring observations of the Crab pulsar
at 327 MHz and 610 MHz with an 85ft telescope in Green Bank, WV showed
a jump in the dispersion by 0.12 cm$^{-3}$ pc. Pulses were seen
simultaneously at both old and new dispersions for a period of
days. In the months before this event faint ghost emission, a replica
of the pulse, was detected with a nearly frequency independent
delay that quadratically diminished to zero. There was also a
curious shift in the phase, a slowdown, at all frequencies
at the time of the dispersion jump. I attribute most of these
phenomena to the perturbing optics of a plasma prism that is
located in the filamentary
interface between the synchrotron nebula and the supernova ejecta
and which crosses the line of sight over a period of months.
The required density, scale length and velocity are reasonable
given detailed HST and previous observations of these filaments.
\end{abstract}

\section{Introduction}
%\noindent
%{\bf 1. Introduction.}
In 1975 the dispersion measure, rotation measure and scattering of the
Crab pulsar displayed an extreme level of activity (Lyne \& Thorne
1975; Isaacman \& Rankin 1977). 
These disturbances in the propagation of radiation from the pulsar were
ascribed to thermal plasma associated with the Crab nebula. In general
the variations of the Crab pulsar's dispersion measure are larger
than that expected for the interstellar medium, and are thus likely the
result of nebular material (Backer \etal 1993).
In recent years the dispersion measure and scattering have been undergoing
a new series of large variations (Backer \& Wong 1996). Column density
variations of 0.1 pc cm$^{-3}$ over time scales of months are seen.
If one places these variations in the filamentary web surrounding the
optical synchrotron nebula, then the likely transverse velocity of the
pulsar-Earth line of sight relative to the plasma $\sim 200$  
km s$^{-1}$.  In this case one can estimate that the characteristic
density of the perturbations 
is $\sim 2500$ e cm$^{-3}$ and the typical transverse length is $\sim 10^{14}$ cm. 
This density is reasonably consistent with density estimates
of the filaments from optical line measurements (Hester \etal 1996, Sankrit \etal 1998)
that have a linear resolution which is several orders of magnitude larger, 
$\sim 10^{16}$ cm. There has also been a unique cluster of glitches in the rotation
of the star during the past few years (Wong, Backer \& Lyne 1999).

In October 1997 amidst this era of large variations of dispersion measure
and other plasma propagation parameters as well as internal ``seismic'' events, 
a dispersion measure ``glitch'' (sudden change in less than one week)
of 0.12 cm$^{-3}$pc was noticed in measurements both at the NRAO Green Bank 
site and at the University of Manchester Jodrell Bank 
site (see Smith \& Lyne in this volume). 
These observatories have small dedicated telescopes for pulsar monitoring.
In fact, during the glitch event pulses were simultaneously detected at both
dispersions. Subsequently Smith \& Lyne found that for about two months prior
to the
dispersion event a faint replica (``ghost'') of the pulsed emission following
the main pulse and interpulse components was
detected in the Jodrell Bank 610-MHz data. Receding ghost components are observed
a few months after the dispersion glitch.
The Green Bank observations confirmed the 610-MHz ghost components and further
provided detections at 327 MHz.
The phase of the
ghost components slowly converged to that of the regular emission a week
before the dispersion glitch. 

While many of the phenomena can be explained by an imperfect plasma
lens passing through the line of sight at the distance of the Crab
nebula filaments,
the occurence of an unusual spindown of the neutron star phase during
the dispersion glitch, and a small
but more conventional spinup glitch of the star two months after the dispersion
glitch 
provide arguments for consideration of plasma propagation in the
vicinity of the star. The ghost of the main pulse and that of the interpulse
are not always identical which adds further confusion to interpretation.
In this brief report, the 327-MHz sequence of events is described. This is followed
by a discussion of the optics that might give rise to these events. A full
report is being submitted for journal publication.

\begin{figure}
\plotone{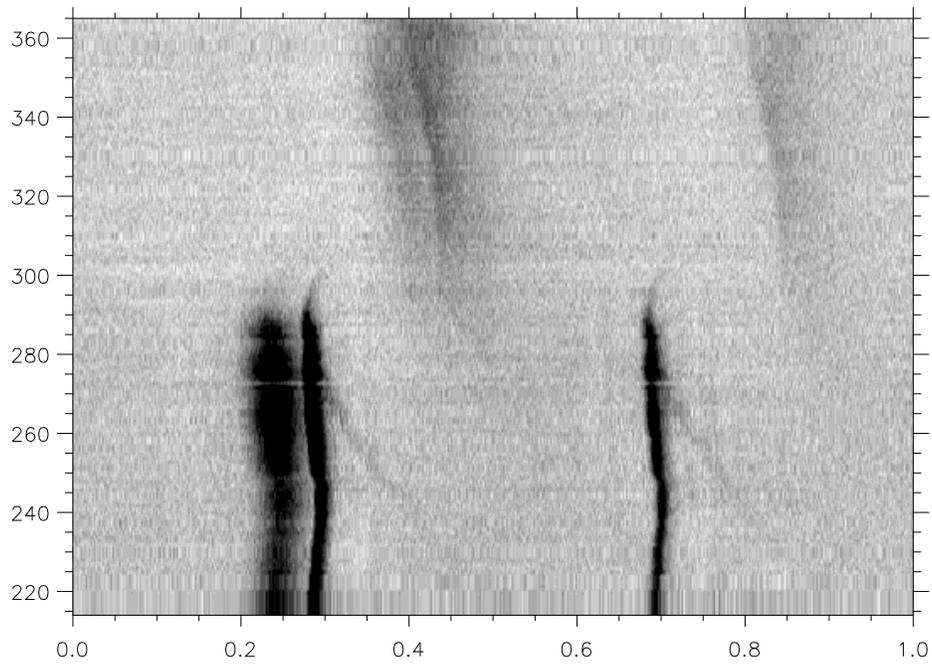}
\caption{
Record of the pulse shapes at 327 MHz during
the interval of 1997 day 225 to 1997 day 365. Pulse fraction is on the
abscissa and day number on the ordinate. The pulse intensity, highly
saturated to show weak emission, is plotted on grey scale. The pulse
consists of precursor, main pulse and interpulse components (from left to 
right). }
\end{figure}

\section{327-MHz event}

%\noindent
%{\bf 2. 327-MHz event.}
I will refer to the pulse profiles in Figure 1 before
and after the dispersion glitch as ``old'' and ``new'' 
respectively. The old pulse peak amplitude is reasonably
steady at the start, then has two peaks centered
on 1997 day 265 and day 275, and then fades away by day 300.
The new pulse is shifted in phase by 0.15 (5 ms) and has amplitudes
which rise steadily from day 290 to day 365. 
The new pulses appear faint owing to a large pulse broadening that I will
attribute to severe multipath propagation in the extra dispersion medium.
The ghost component of the main pulse appears at phase 0.4 around day 250
and moves along a quadratic path toward the main pulse at phase 0.28 around
day 275. The ghost of the interpulse follows a similar path.
The peak intensities are typically 1-2\% times that of the corresponding
pulse components.
Note that during days 285 to 299 there is emission present simultaneously
from old and new pulses.  I interpret this
in terms of two propagation paths from the pulsar in the next section.

Extrapolation of the old pulse phase to day 320 leads to an
offset between new and old pulses of 5.1 ms at 327 MHz. The multipath
propagation delay from both new and old pulses have been removed in this
estimate. The corresponding
number at 610 MHz is 2.1 ms. These two delays are {\it not} consistent with the
cold plasma, quadratic dispersion law. There appears to be a 0.9-ms
achromatic delay. The source of this delay is actually just 
detectable in the fading old pulse emission: in Figure 1
the peak of the main and interpulse is rapidly drifting to the right
during days 290 to 299. Most of this drift is achromatic; {\it i.e.,}
it is seen at both 327 MHz and 610 MHz. Achromatic variations of pulse
arrival time are most naturally interpreted as seismic events internal
to the neutron star. Thus just at the time of
this dispersion glitch with emission seen along two lines of sight
simultaneously, there appears to be a spindown of the pulsar of
rather large amplitude -- large with respect to the timing noise in
this pulsar which is characterized by a random walk in frequency.
I will proceed with discussing the ghost component and dispersion glitch
in terms of propagation through a plasma prism in the filamentary
interface of the Crab nebula, and leave the ``coincidence'' of an
internal effect aside. A final remark here, in admittedly weak
support of ignoring the nominally internal event, is that the pulsar
has been extremely active with many spin glitches over the 1996-1998
interval.

\begin{figure}
\plotone{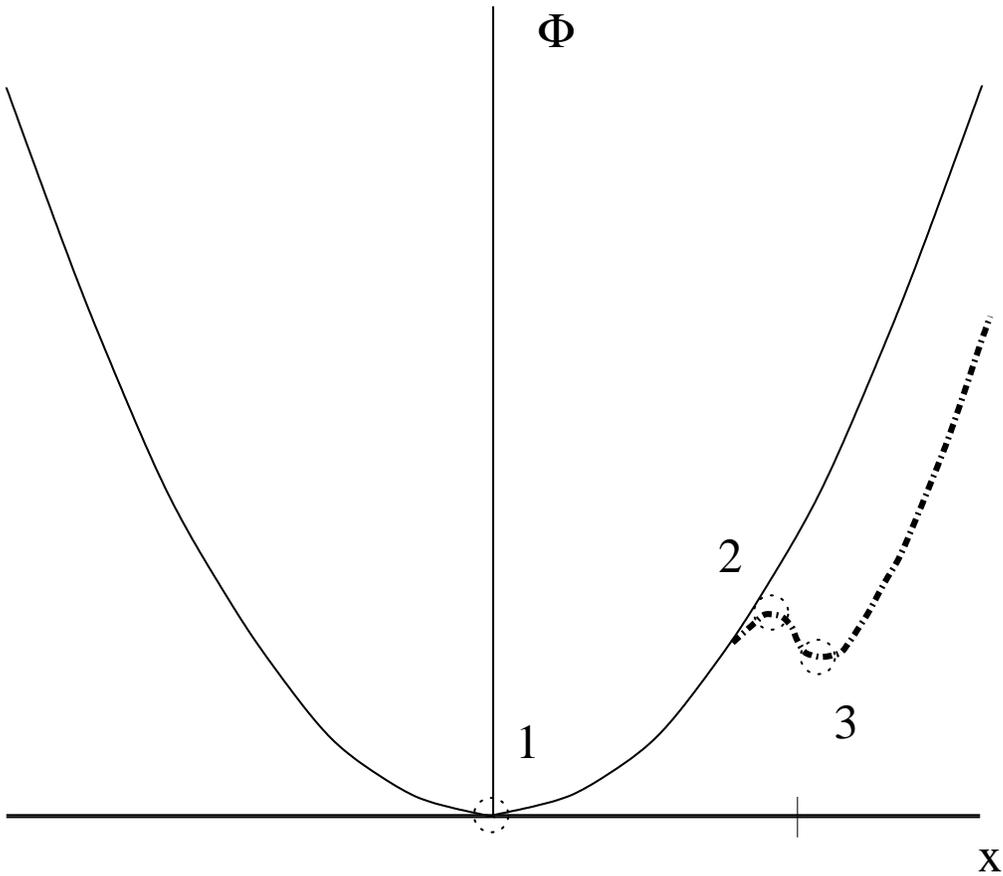}
\caption{Wave phase through the plasma prism
including quadratic Fresnel bowl term and the leading
edge of the prism. Stationary phase points are
enumerated which result along the geometric
line of sight (1) and at the edge of the prism
(2,3).}
\end{figure}

\section{Plasma Prism Model}

%\noindent
%{\bf 3. Plasma Prism Model.}
The dispersion glitch and the simultaneous appearance of radiation
along nominal dispersion (old) and extra dispersion (new) paths
can readily be explained by a wedge of plasma moving into the
line of sight. The extra dispersion largely disappears after
250 days. The start of the steady decline can be seen in the 
drifting location of the new pulse in Figure 1 between days 300 and 365.
Thus the dispersion changes can be explained as a plasma prism
passing through the line of sight and oriented so that
one observes a fairly abrupt change of dispersion. The refractive
property of a simple uniform density plasma prism, with its
cold plasma index of refraction less than unity, is such
as to bend the new pulse into the line of sight and allow simultaneous
observation of the old and new pulse.  From
the observed gradient in dispersion I estimate a refraction
angle of 1 $\mu$as at 327 MHz. Refraction of this magnitude 
at a point 2 pc along the line of sight from  the pulsar
to Earth would lead to a small time delay of 0.2 ms at 327 MHz. Of
course this {\it geometrically} delayed path would also have
the new dispersion delay.

Figure 2 presents the optics of this plasma prism model more
explicitly by taking a cut of the total wave phase as the wave
exits the plasma prism. The wave phase consists of a geometric
part, a cut through a quadratic `Fresnel bowl', and the plasma
prism part shown to the right of the line of sight. What one
observes at any instant is signal from the stationary phase
points. With a plasma prism, as with a gravitational lens, one
sees an odd number of signal paths, in this case 3. Path 1 is
the unperturbed geometric line of sight. Path 2 is on the
leading edge of the prism and path 3 from within the prism.
Thus path 2 would have approximately the same dispersion 
as path 1 while 3 would have the excess dispersion of the 
prism. The flux one observes from each stationary point depends
on the size of the stationary phase patch, or more explicitly
the second derivative of the phase ({\it e.g.} Clegg et al. 1998). 
For the unperturbed line
of sight, the size is that of the first Fresnel zone.
As the line of sight moves toward the prism's edge, the location
of 2 and 3 changes roughly linearly, and the excess delay along
this path will change approximately quadratically. Further
this geometric path delay effect is approximately achromatic. Thus we
can identify point 2 with the ghost component and point 3
with the simultaneous appearance of new and dispersed pulse.
The signal from point 3 comes through the body of the prism
which, as stated above, has the further property of excess 
multipath propagation. This effect can lead to the difficulty
of seeing the third pulse during days when the ghost pulse
is seen (245-275).

\section{Conclusion}

%\noindent
%{\bf Conclusion}
I have presented the 327-MHz Crab pulsar data from late 1997. These
show a number of events -- ghost components and dispersion glitch and
spindown. The principal features can be explained by propagation through
a plasma prism that takes several months to drift through the line
of sight. The physical parameters of the prism -- size of $3\times 10^{14}$ cm
and density of 1200 e cm$^{-3}$ -- are reasonably consistent
with those determined by forbidden line observations of the filamentary
gas surrounding the optical synchrotron nebula. The density is possibly high
given the small length scale in relation to the filamentary structure
discussed by Hester \etal 1996 and Sankrit \etal 1998. The density can
be reduced by choosing a sheet or pencil geometry that extends into
the plane of the sky by larger scale than what is detected in the transverse
direction (see similar argument for interstellar tiny HI structures in
Heiles 1997).
\bigskip
\bigskip
\noindent
{\bf Acknowledgements} I thank the NRAO staff at the Green Bank site who
have helped to maintain the pulsar monitoring telescope program there,
Graham Smith and Andrew Lyne for many discussions on these still puzzling
events, and Berkeley students Tony Wong and Jay Valanju who assisted
in data analysis.

\end{document}